# Electronic Bound States in the Continuum in a 2D Metal

**Authors:** Morgan Thinel[1,2], Simon Turkel[1,3], Sebastian E. Rossi[1], Christie S. Koay[2], Daniel G. Chica[2], Xiong Huang[1], Madisen Holbrook[1], Aravind Devarakonda[4], Luca M. Nashabeh[1], Alexandru B. Georgescu[5*], Xavier Roy[2*], Xiaoyang Zhu[2*], Abhay N. Pasupathy[1,3*], and Raquel Queiroz[1*]

**Affiliations:**

[1] Department of Physics, Columbia University; New York, NY, USA
[2] Department of Chemistry, Columbia University; New York, NY, USA
[3] Condensed Matter Physics and Materials Science Division, Brookhaven National Laboratory; Upton, NY, USA
[4] Department of Applied Physics and Applied Mathematics, Columbia University; New York, NY, USA
[5] Department of Chemistry, Indiana University; Bloomington, IN, USA
\* Corresponding authors. Email: georgesc@iu.edu, xr2114@columbia.edu, xz2324@columbia.edu, apn2108@columbia.edu, rq2179@columbia.edu

**Abstract:** Bound states in the continuum (BICs) are quantum states that remain localized despite existing within a continuum of extended, delocalized states. They defy conventional wave theories and could be instrumental for quantum technologies that rely on the precise control of quantum states. While optical BICs have been realized in photonic systems, achieving electronic bound states in a metallic background remains an ongoing challenge. Here, we observe two defect states that remain localized within the metallic continuum of $Pd_5AlI_2$, a two-dimensional van der Waals metal. The emergence of these states is a manifestation of the hopping interference in the $Pd_5AlI_2$ lattice. This interference results in a (quasi) flat band and spatially localized eigenstates that are orthogonal to the metallic continuum thus avoiding hybridization with extended states.

**Main Text:** Solutions to wave equations usually feature a spectral separation between extended continuum states and spatially localized discrete frequencies. In the early days of quantum mechanics, however, von Neumann and Wigner showed that specific potentials in the Schrödinger equation could result in spatially localized wavefunctions with energies within the continuous spectrum of extended states (*1*). These bound states in the continuum (BICs) have been shown to emerge from diffractive interference, making them a generic property of wave equations, including the Maxwell equations in acoustics and photonics (*2*). BICs have been experimentally realized in photonic systems, where nanofabrication tools enable customization of the energies and symmetries of photonic modes (*3-5*). Applications such as chiral lasing (*6-9*), enhanced nonlinear frequency conversion (*10-12*), and the mitigation of radiation losses in waveguide devices have emerged from this research (*13-15*). In general, BICs are designed via fine parameter tuning, which enforces localized destructive interference of extended waves (*16, 17*). Other approaches have made use of point defects in photonic lattices to stabilize BICs protected by lattice symmetries (*18, 19*). These advancements motivate the search for analogous electronic bound states within a metallic continuum in condensed matter systems.

Bound states around defects are ubiquitous in semiconductors and insulators, existing in a spectral gap between bands (*20-24*). The localization of these states is directly tied to the presence of the bandgap, allowing the defect states to avoid hybridization with continuum states. In metals, however, dilute point defects behave as scattering centers for extended states, which only weakly perturbs them (*25, 26*). Many-body interactions such as the Kondo effect (*27*) can give rise to localized resonances (*28-31*) but they are typically found near the Fermi level rather than deep within the metallic band. While a few studies have proposed the existence of electronic BICs, they lack the spatial resolution to demonstrate their strongly localized nature (*32-34*). By analogy to optical BICs, modulating lattice hopping parameters may be a promising avenue to realize electronic BICs (*35*). In this work, we report a general mechanism that produces robust BICs in lattices exhibiting flat bands due to frustrated electron hopping between adjacent lattice sites. This frustration occurs in bipartite lattices with an unequal number of sublattice sites like the Lieb lattice (*36*), as well as in line graph lattices such as the kagome (*37*) and dice (*38*) lattices. Frustrated hopping can also result from the interplay of lattice symmetry and orbital composition, such as in the decorated checkerboard lattice (*39, 40*). In all these lattices, dispersive bands coexist with highly degenerate flat bands, in which the extended Bloch states combine into spatially localized eigenstates known as compact localized states (CLSs) (*41-43*). The formation of electronic BICs can be understood within this framework (Fig. 1A): a lattice defect shifts the chemical potential of a nearby CLS into the continuum of dispersive bands. Due to the localized nature of the eigenstates, the more distant CLSs remain largely unaffected while the continuum states are only slightly perturbed. Importantly the orthogonality between the shifted CLS and the dispersive states is preserved such that the CLS remains spatially localized and does not hybridize with the continuum, forming a BIC. This mechanism is intrinsic to materials hosting nontrivial flat bands (*44, 45*), enabling the formation of electronic BICs through local vacancies or impurities without requiring fine-tuning model parameters.

Here we use scanning tunneling microscopy and spectroscopy (STM/S) to directly characterize the local density of states (LDOS) of electronic BICs in $Pd_5AlI_2$, a van der Waals (vdW) metal with a decorated checkerboard lattice that gives rise to frustrated electron hopping (*46*). The tetragonal crystal structure of $Pd_5AlI_2$ (space group *I4/mmm*) consists of a stack of vdW slabs, each five atoms thick. These slabs can be peeled off the crystal and isolated as air stable flakes through mechanical exfoliation. Each intermetallic slab is composed of a PdAl checkerboard lattice sandwiched between two square planes of $Pd_2$, capped with iodine atoms (Fig. 1B). Tight binding modeling of the PdAl layer reveals it forms a bipartite lattice with frustrated hopping between orthogonal Pd $d_{xz}$ and $d_{yz}$ orbitals, mediated by Al $p_z$ orbitals (Fig. 1C) (*46*). The electronic structure of this Lieb-like lattice, termed decorated checkerboard, has been measured in photoemission experiments and features linearly dispersing bands at high symmetry points (M and Γ) which are intersected by a flat band (*46*). Figure 1C shows a CLS of this flat band mapped onto the PdAl inner-layer of $Pd_5AlI_2$. Figure 1D shows atomic resolution STM topography of the surface iodine atoms of a cleaved $Pd_5AlI_2$ crystal, measured at negative sample bias. The image features a square lattice with an in-plane lattice constant $a = 4.05$Å (Fig. S1), in agreement with single crystal X-ray diffraction (*47, 48*). Three types of point defects, labeled as i, ii, and iii in Fig. 1D, are observed in relatively high density (~ $2.3 \times 10^{12}$ cm$^{-2}$) across the surface (Fig. S2). The tunneling spectrum of $Pd_5AlI_2$ shows a non-zero DOS at all bias voltages (purple spectrum, Fig. 1E), consistent with the metallicity of the material. Remarkably, STS performed on type ii defects reveal two distinct peaks in the LDOS centered at +475 and +620 mV within the continuous metallic bands (green spectrum, Fig. 1E).

We identify this defect type as Pd vacancies in the central decorated checkerboard PdAl plane by comparing experimental STM topography and DFT simulations at different sample biases (Fig. 2A, B). Spectroscopic imaging over the region shown in Fig. 2C reveals a four-lobed pattern around the Pd vacancy at +475mV, which is reminiscent of an atomic $d$-orbital (Fig. 2D). The state at +620 mV appears as a circular pattern that resembles an atomic $s$-orbital (Fig. 2E). By measuring point spectra along the blue line in Fig. 2A, we demonstrate that the states are confined to a sub-nanometer spatial range (Fig. 2F and DFT calculations in Fig. S3), indicating a remarkable localization of both states. The presence of discrete states within the metallic band, together with their spatial localization suggest they are electronically isolated from the metallic continuum, akin to optical BICs isolated from the photonic continuum.

We turn to the frustrated hopping band structure of the PdAl layer containing this defect to explain the localization effect. The band structure for an idealized decorated checkerboard model with no next-nearest-neighbour hopping is shown in Fig. 3A. In Fig. 3B, we use the simplified three-band model describing Pd $d_{xz}$ and $d_{yz}$ orbitals and Al $p_z$ orbitals to compute the DOS spectrum for a 41×41 supercell of PdAl, with a single Pd atom removed from the supercell. The spectrum around the vacancy shows excellent agreement with experiment, where two states emerge a few hundreds of meV above the Fermi level ($E_F$). These eigenstates are formed by symmetric CLS states in a 3x3 grid surrounding the Pd defect (Fig. S4). The relative phases of the neighboring orbitals forming the localized states are constrained by the $C_4$ symmetry of the system about the vacancy, resulting in four possible eigenstates that we label by their eigenvalue under the $C_4$ symmetry operator: +/−1 or +/− $i$. When examining the symmetry of the localized states (Fig. 3B), we find that the lower energy state at +400 meV corresponds to the −1 eigenstate while the state at +700 meV corresponds to the +$i$ eigenstate. An additional on-site potential is necessary for the +$i$ state to emerge, which we attribute to $p$ orbitals from the iodine atoms positioned above and below the PdAl checkerboard (Fig. S5). These $p$ orbitals are affected by the absence of the Pd orbitals in the plane. By mapping the nodal lines (Fig. 3C, D), defined by π-phase shifts of the wavefunctions, we find that the −1 state features angular nodal planes, forming a four-lobe structure, while the +$i$ state has a radial node, resulting in a circular structure. These theoretical results are consistent with experimental spectroscopic imaging (Fig. 3E, F).

To understand whether the perfectly flat band of the ideal decorated checkerboard lattice is a necessary condition to host point defect BICs in Pd$_5$AlI$_2$, we artificially introduce next-nearest neighbor hopping to our model, which adds dispersion to the flat band. This transformation makes the CLSs not exact eigenstates of the system, and therefore, the defect state can hybridize with the continuum of extended states. However, the exponential localization of the BICs is robust for a large parametric range of next-nearest neighbor hopping and band dispersion, including realistic parameters that match the experimental band structure of $Pd_5AlI_2$ (Fig. S6) (*46*). This confirms that localization is driven by the topologically nontrivial character of the flat band, which emerges from the interference of neighboring orbitals and does not find an analog in isolated orbitals (*49-51*). The atom-like orbitals observed in Fig. 2D, E are indeed larger than atomic orbitals.

These electronic BICs resemble atomic orbitals of different symmetries at different energies. Like conventional atomic orbitals, they can bind when in proximity, forming molecular bound states. However, unlike atomic orbitals, their energy splitting into bonding and antibonding orbitals does not need to respect conventional rules. We perform spectroscopic imaging over an area containing two defects separated by a single atomic lattice constant (Fig. 4A). Spatially-averaged spectroscopy over this dimer reveals four bound states, corresponding to the splitting of each of the two BICs associated with a single vacancy (Fig. 4B). This splitting is reminiscent of bonding and anti-bonding energy levels that form in homonuclear diatomic molecules. Consistent with the expectation for bonding in diatomic molecules, spectroscopic imaging of divacancy states of the four-lobed BIC shows a bonding orbital at +440 mV and an antibonding orbital at a higher energy of +510 mV (Fig. 4C, D). In contradistinction, the hybridization of the circular states creates a node between the defects, similar to an antibonding molecular orbital, at +590 mV (Fig. 4E). The higher energy state at +650 mV exhibits a high density of states between the defects, resembling a bonding molecular orbital (Fig. 4F). Spectroscopic imaging over an area containing a dimer and an isolated vacancy confirms this energy distribution (Fig. S7). When Pd vacancies are separated by two or more atomic lattice constants, inter-defect hybridization is not observed, further supporting the strongly localized nature of the BICs (Fig. S8). We can describe these dimerized states within the PdAl tight binding model as linear combinations of two single-defect eigenstates (Fig. 4G-J). In diatomic molecules, bonding orbitals result from favorable phase interactions in the linear combination of atomic orbitals, which make up the molecular orbital. Indeed, the relative phase of the BIC with respect to its neighbour—i.e. the sign of the linear combination—determines the energy of the bonding and antibonding states (Fig. S9). In particular, when two defect states are out of phase with each other, the interaction is favourable and the divacancy state is lower in energy. The dimer states formed by combining two in-phase defect states are higher in energy.

Using STM/S, we have demonstrated that electronic bound states form around vacancies within the metallic continuum of $Pd_5AlI_2$. This phenomenon arises from electron hopping interference on the decorated checkerboard of $Pd_5AlI_2$, leading to spatially localized eigenstates that are orthogonal to the continuum and immune to hybridization. Our results demonstrate that point defects in $Pd_5AlI_2$ can host a condensed matter analog of optical BICs. More broadly, our results point to a general mechanism to create BICs in other 2D frustrated hopping lattices, such as the Lieb and Kagome lattices, which can naturally occur or be artificially designed in various condensed matter systems (*52-55*). By analogy to the infinite lifetime of photonic BICs, electronic BICs could support lossless transport for charge carriers at the BIC energies. Furthermore, we have demonstrated that defect dimerization controls the energy of these electronic states, which can be used to demonstrate their unique topologically nontrivial character. Motivated by recent advances in defect engineering demonstrating the assembly of atomically precise vacancy lattices in vdW materials such as semimetallic $PtSe_2$ (*56*), we envision $Pd_5AlI_2$ could provide a platform for the controllable synthesis of electronic BIC lattices that may host topological properties. Such atomically precise topologically nontrivial lattices could be useful for applications in catalysis and energy storage (*57*). Additionally, introducing magnetic impurities at the Pd vacancy site could utilize the Kondo effect to pin electronic BICs to the Fermi level and influence bulk transport properties. Finally, the vdW structure of $Pd_5AlI_2$ provides a path for interfacing with other materials through mechanical exfoliation, stacking and twisting. Overall, the thermodynamic and kinetic control of defect density and migration, in conjunction with conventional vdW material manipulation, make $Pd_5AlI_2$ a promising candidate for highly tunable localized states that may be useful for topological quantum chemistry as well as topological quantum computation devices.

**Acknowledgments:** The DFT calculations shown in this work have been performed on the Big Red 200 and Quartz supercomputer clusters at Indiana University.

**Funding:** This work on localized states in frustrated hopping metals was primarily supported by the National Science Foundation through the Columbia University Materials Research Science and Engineering Center (MRSEC) on Precision-Assembled Quantum Materials DMR-2011738 (R.Q., A.N.P., X.-Y.Z. and X.R.). STM measurements were partially supported by the Air Force Office of Scientific Research under award numbers FA9550-22-1-0389 (X.-Y.Z. and X.R.) and FA9550-21-1-0378 (A.N.P.). Synthetic work was partially supported as part of the Programmable Quantum Materials, an Energy Frontier Research Center, funded by the US Department of Energy (DOE), Office of Science, Basic Energy Sciences, under award DE-SC0019443 (X.R., A.N.P., X.-Y.Z.). Theoretical work at Columbia University was partially supported by the National Science Foundation under award number DMR-2340394 (R.Q.). DFT calculations at Indiana University were supported in part by Lilly Endowment, Inc., through its support for the Indiana University Pervasive Technology Institute, as well as by Indiana University startup funds (A.B.G).

**Author contributions:**

STM experiments: M.T., S.T.

Synthesis: C.S.K., D.G.C., X.R.

Tight-Binding Calculations: S.R., M.T., L.M.N., R.Q.

DFT Calculations: A.B.G.

Supervision: R.Q., X.R., X.-Y.Z., A.N.P.

Resources (code): A.D.

Writing – original draft: M.T.

Writing – review & editing: all authors

**Competing interests:** none

**Data and materials availability:** available upon request

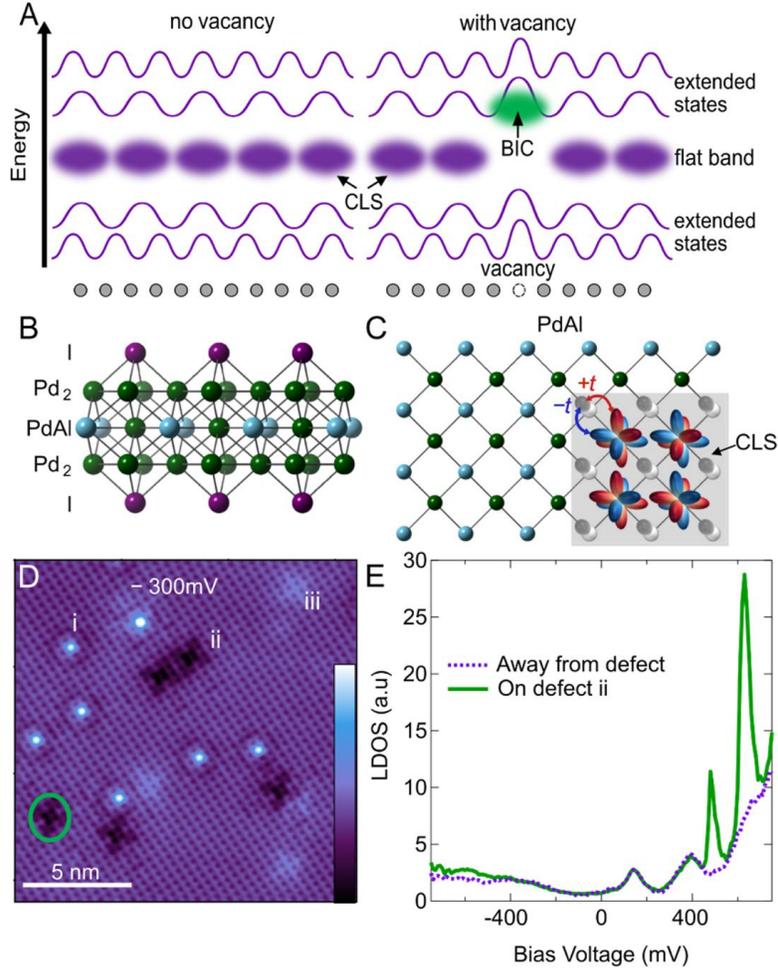

**Fig. 1. Bound States on Defects in $Pd_5AlI_2$. A.** Schematic of extended states in a metallic continuum and compact localized states of a flat band. A flat band eigenstate that is pushed into the continuum from the local potential of a defect remains localized since it is orthogonal to the continuum states. **B.** Crystal structure of a $Pd_5AlI_2$ vdW slab, viewed along the *ab* plane. **C.** Top view of the central PdAl checkerboard plane with an overlay showing an atomic orbital representation of the CLS (grey square). Frustrated hopping occurs between orthogonal Pd $d_{xz}$ and $d_{yz}$ orbitals, mediated by Al $p_z$ orbital. Blue and red arrows indicate the in-phase and out-of-phase hopping, respectively. **D.** Constant current STM topography image of a $16 \times 16$ nm$^2$ area of $Pd_5AlI_2$ (Setpoint: −300 mV, 50 pA) showing surface iodine atoms and three types of point defects. **E.** Average tunneling spectrum (Setpoint: 150 mV, 30 pA; Modulation: 25 mV, 927 Hz) measured on a defect-free region (purple dashed line) and on a type ii defect (green). Two bound states in the continuum emerge on defect ii.

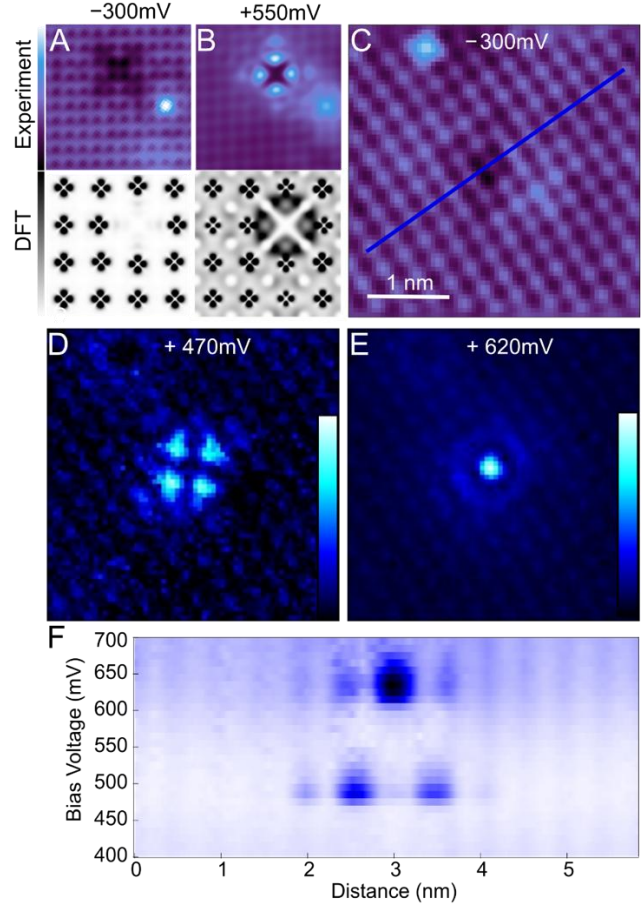

**Fig. 2. Spectroscopic Structure of Bound States on Pd Vacancies. A, B.** Experimental (purple) and DFT simulated (greyscale) STM topography of $Pd_5AlI_2$ areas containing a Pd vacancy at (A) negative (−300 mV, 50 pA) and (B) positive (+550 mV, 50 pA) sample bias. **C.** 5 × 5 $nm^2$ STM topography of a region containing all three types of defects, where STS maps were acquired. **D, E.** Real space LDOS maps at bias voltages corresponding to the defect-induced peaks in LDOS (**D** = +470 mV, **E** = +620 mV). Color scale indicates LDOS. Strong features in the DOS are apparent only around the type ii defect and have different orbital-like symmetries at the two quantized energy levels. **F.** Line spectroscopy across the region delimited by a blue line in panel C. The defect states decay within 1 nm from the centre of the defect. Data in panels C-F were acquired with a feedback setpoint of −300 mV, 50 pA. Spectra were acquired with a lock-in modulation of 10 mV, 927 Hz.

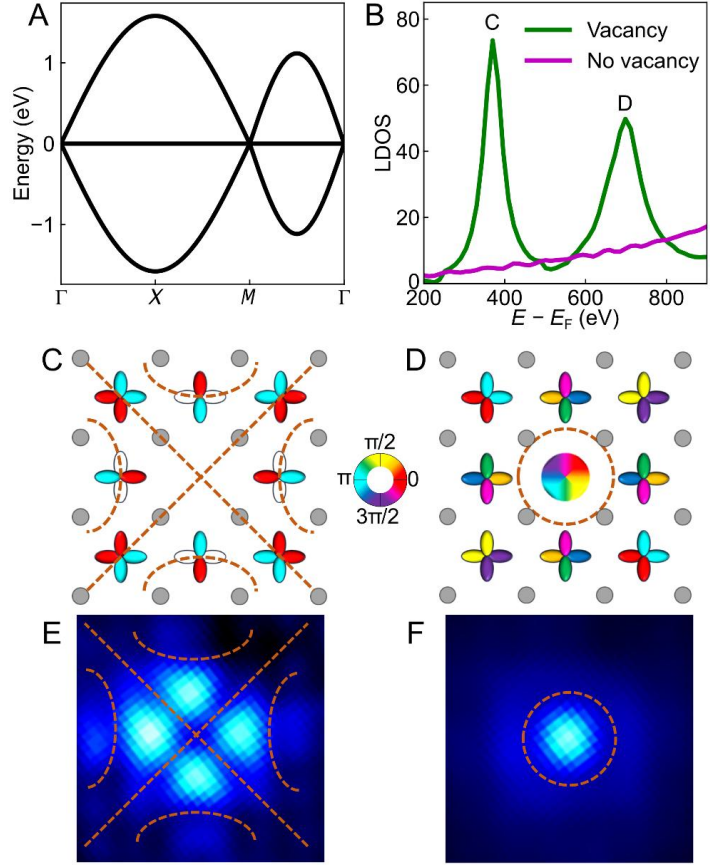

**Fig. 3. Nodal Structures in a Frustrated Hopping Model. A.** Band structure of an idealized decorated checkerboard model where hopping interference leads to a flat band intersected by two linearly dispersing bands. **B.** DOS spectra from a simulated 41 × 41 supercell of decorated checkerboard. The spectrum for the 3 × 3 supercell around a Pd vacancy (green) reveals two peaks that are absent in a 3 × 3 supercell without the vacancy (purple). **C, D.** Schematic of two eigenstates of the 41 × 41 supercell containing a single Pd vacancy. Colour wheel indicates the relative phase of Pd $d$ orbitals. Nodal lines (orange dashed lines) emerge where there is a $\pi$-phase shift of the wavefunction. The state in C emerges because of an additional $\sigma_y$ transformation from an on-site potential from iodine $p_x+ip_y$ orbitals (rainbow circle, Fig. S5). These eigenstates are combinations of flat band CLSs, and thus are also eigenstates of the flat band. They are orthogonal to high energy dispersive states and remain localized. **E, F.** Nodal lines from the tight binding model overlaid on the experimental spectroscopic maps shown in Fig. 2, showing excellent qualitative agreement.

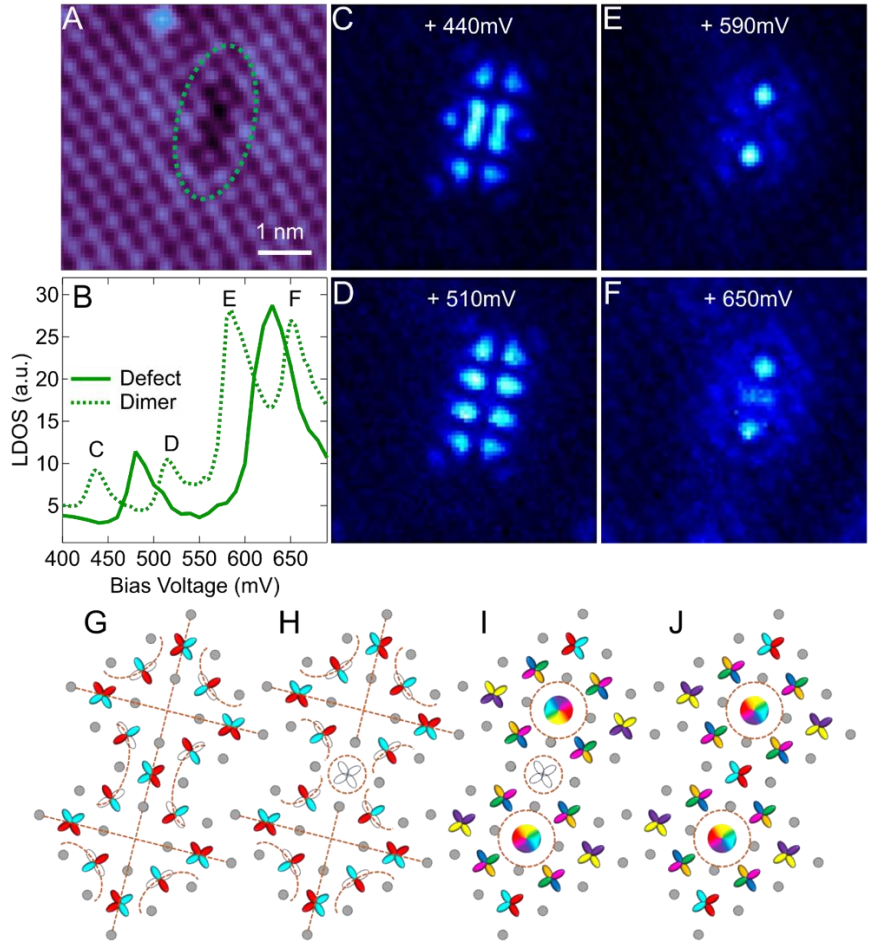

**Fig. 4. Dimers of Electronic BICs. A.** STM topography of the region containing a divacancy where spectroscopic maps were acquired. **B.** Spatially averaged tunneling spectrum over a defect dimer (dashed oval) overlaid with the single defect spectrum from Fig. 1D. **C-F.** Real space LDOS maps at bias voltages corresponding to the dimer bound state peaks in panel B. **G-J.** Schematic of dimer eigenstates formed by combining two vacancy eigenstates. G and I are combinations where there is a $\pi$-phase relationship between the two vacancy states. H and J are combinations of two in-phase defect states (Fig. S9). There is qualitative agreement between nodal lines defined in G-J and the shape of the experimental data. Data were acquired with a feedback setpoint of −400 mV, 50 pA and lock-in modulation of 5 mV, 1127 Hz.

# Supplementary Materials for

## Electronic Bound States in the Continuum in a 2D Metal

**Authors:** Morgan Thinel[1,2], Simon Turkel[1,3], Sebastian E. Rossi[1], Christie S. Koay[2], Daniel G. Chica[2], Xiong Huang[1], Madisen Holbrook[1], Aravind Devarakonda[5], Alexandru B. Georgescu[4*], Xavier Roy[2*], Xiaoyang Zhu[2*], Abhay N. Pasupathy[1,3*], and Raquel Queiroz[1*]

*Corresponding author. Email: georgesc@iu.edu, xr2114@columbia.edu, xz2324@columbia.edu, apn2108@columbia.edu, rq2179@columbia.edu

**The PDF file includes:**

    Materials and Methods
    Supplementary Text
    Figs. S1 to S9
    References

**Materials and Methods**

Synthesis of $Pd_5AlI_2$

$Pd_5AlI_2$ crystals were synthesized via a direct combination reaction of a slightly off stoichiometric amount of the elements. Briefly, palladium powder (100 mg, 0.940 mmol, Fisher Scientific, 99.95%), aluminum chunks (7 mg, 0.259 mmol, Strem Chemicals, 99+%), and iodine chunks, (47.7 mg, 0.188 mmol, Aldrich, 99.99+%) were loaded into a fused silica tube with a O.D. and I.D. of 9 mm and 7 mm, respectively. The tube was sealed to a length of 15 cm under $\approx$ 30 mTorr of Ar while under liquid nitrogen to reduce the loss of iodine. The tube was heated to 600°C in 6 hours, annealed for 84 hours, and then water quenched to ambient temperature. The compound formed as shiny square metallic crystals with lateral dimensions of 1-10 mm and thicknesses of <100 μm. A detailed account of the synthesis can be found in ref #46.

Scanning tunneling microscopy/spectroscopy

All STM/STS measurements were performed in a home-built STM at a sample stage temperature of ~ 7 K using an electrochemically etched tungsten tip. Prior to each sample approach the tip was calibrated against the Au (111) surface state. Samples were cleaved in UHV prior to measurement. Spectroscopy was recorded using a lock-in amplifier to measure the differential conductance dI/dV. Setpoint and lock-in modulation information is found in the Fig. captions.

DFT

Density functional theory calculations were performed using the Quantum Espresso code version 6.8, using ultrasoft pseudopotentials, and the PBE version of the GGA exchange correlation formalism. To simulate a monolayer with a Pd vacancy defect, a 4×4×1 unit cell was simulated allowing 12 Angstroms of vacuum along the $z$ direction. The lattice vectors were fully relaxed along the $xy$ plane but not allowed to relax along $z$ to maintain the vacuum, with the atomic positions free to move in all 3 directions. The energy cutoff used was 544eV, and a $k$-mesh of 2×2×1.

**Supplementary Text**

Characterization of Iodine Lattice and Point Defects

In Fig. S1, the STM topography of $Pd_5AlI_2$ at negative sample bias is further investigated. On one hand, we identify the square lattice that appears in STM topography as corresponding to the lattice of iodine atoms in the I-plane of $Pd_5AlI_2$ (Fig. 1 of main text for structure). The lattice constant of 0.405nm established from diffraction can be compared to measurements of the fast-Fourier-transform of STM topography images. We consistently obtain a similar value when measuring the distance between Bragg wavevectors $q_{Bragg}$ obtained from the 2D-FFT of STM topography (S1B, D). Through real space measurements, we can identify the lattice locations of defects of interest. Fig. S1A,C show that the dark defects are found on the same lattice site as the iodine atoms, whereas the bright defects are found in between. In the PdAl plane, these lattice sites correspond to Pd (dark) and Al (bright) respectively. This is consistent with our interpretation of the defect of interest being a Pd vacancy.

In Fig. S2, we show a large (55×55 nm$^2$) field-of-view (FOV) STM topography of two different areas of the crystal. Data were acquired with a positive (A) or negative (B) sample bias. We proceed to count defects in 5 different 55x55 nm$^2$. We obtain an average defect density between 2-2.5×10 cm$^{-2}$ for each type of defect. At positive bias, defects of interest show a four-lobed structure as shown in Fig. 2 of the main text.

DFT Calculations on Pd Vacancies

In Fig. S3, we simulate STM topography and spectroscopy for a 4×4 supercell of $Pd_5AlI_2$ with a single Pd vacancy. We simulate STM topography at negative (A) and positive (B) sample biases and investigate the appearance of the PdAl plane. Consistent with experimental STM topography in Fig. 2, only positive sample bias topography yields a localized four-lobed shape around the Pd vacancy. In Fig. S3C, we calculated the band structure with and without a Pd vacancy. The emergence of a new flat band in the presence of a vacancy supports the presence of a localized state emerging on this vacancy. In S3D, we calculated the partial density of states (PDOS) for a location far from the vacancy and a location on the vacancy. We observe a peak in the DOS spectrum at around 0.55-0.6 eV above the Fermi level, which is consistent with the experimental observation of a peak centered around 0.47 eV above $E_f$ in the experiment (see Fig. 1 main text).

Details of Tight-Binding Calculations of Pd Vacancies

In Fig. S4, we describe the frustrated hopping model used to rationalize the presence of BICs. The flat band eigenstate is formed through interference between distinct orbitals in the PdAl layer of $Pd_5AlI_2$ (Fig. S4A). The phase relationship between the central Pd and corner Al sites leads to the formation of a compact localized state (Fig. S4B) where destructive interference of electron hopping leads to a flat band. These compact localized states can be combined to form vacancy eigenstates, where the Pd vacancy perturbs the four compact localized states to which it belongs. The resulting 3×3 states are also eigenstates of the frustrated hopping flat band, since they are composed of linear combinations of the four CLSs. The orthogonality between flat and dispersive states (Fig. 1 main text) means these states must be orthogonal to the dispersive states that exist in the metallic continuum. As such, when a defect potential pushes these states to some energy outside the flat band, they remain localized to the defect location and isolated from the continuum. They are electronic bound states in the continuum.

In Fig. 3D of the main text, a circle on the defect location represents an additional $\sigma_y$ transformation that is necessary for the $+i$ state to emerge in our model. In Fig. S5, we show the structure of $Pd_5AlI_2$ with a Pd defect in the PdAl plane. The iodine atoms above and below the defect have $p_x+ip_y$ and $p_x-ip_y$ orbitals which contribute to the on-site potential of the defect. The simulations in Fig. S6 are done with the following parameters:

Nearest-neighbour hopping (between Pd and Al): $t$=0.395eV
Next-nearest neigbour hopping: $t_{nnn}$=0 to 3.5
$t_{nnn}$ between Pd orbitals: $t_1$=0.15×$t_{nnn}$
$t_{nnn}$ between Al orbitals: $t_2$=0.038×$t_{nnn}$

We note that within this framework, a $t_{nnn}$ value of 1 corresponds to a band structure that is representative of the experimental flat band resolved in ref. 46.
The presence of the defect affects the atoms around it, so we add the following potentials to the 3×3 supercell around the vacancy:

Scalar potential: $U$=0.25V
$\sigma_y$ potential: $m$=1V, where a transformation $m*\sigma_y$ is applied to the Pd $d$ orbitals around the vacancy.

The described Hamiltonian was used to generate a 41×41 supercell, where each cell contains the two Pd $d$-orbitals at the center and Al $p$-orbitals at the corners. The central Pd orbitals of one unit cell was removed by forcing the hopping parameter to zero at the vacancy location. The band structure in Fig. 1 and spectrum in Fig. 3 of the main text were calculated with $t_{nnn}=0$ and by taking the density-of-states within 3-unit cells of the vacancy for the vacancy spectrum and the density-of-states of a 3×3 cell with no vacancy for the no vacancy spectrum.

To generate the localized eigenstates shown in Fig. S6, we computed the DOS spectrum for the vacancy selected values of $t_{nnn}$. For each spectrum, we calculated the inverse participation ratio (IPR=$1/\psi^4$) of the eigenstates closest in energy to the peaks in the spectrum. We then plotted the state with the lowest IPR (most-localized state) for each value of $t_{nnn}$ (selected values of $t_{nnn}$ shown in Fig. S6E-H). The IPR and energy of each eigenstate are labelled in the Fig.. We note that with a constant on-site potential (0.25eV), the eigenstate shifts in energy as the flat band is moved up in energy due to the increased $t_{nnn}$. We hypothesize that the threshold $t_{nnn}$ where the BICs become trivially delocalized defect resonances is when the energy of the eigenstate begins to overlap with the flat band and can hybridize with it.

In Fig. S6, we verify that BICs are a consequence of the frustrated hopping lattice model formed by the PdAl plane in Pd$_5$AlI$_2$. In Fig. S6A-A, we calculate the band structures for our 3-band model with increasing values of the next-nearest-neighbour hopping parameter $t_{nnn}$. Fig. S6A-D are the band structures for values of $t_{nnn}$=0.5, 2.0, 3.0, and 3.5 respectively. Fig. S6E-H show the wavefunction of the four-lobed localized eigenstate around a vacancy in a 41×41 supercell for increasing values of $t_{nnn}$. As the flat band model is destroyed via artificially adding $t_{nnn}$ the localized BIC becomes a trivially delocalized metallic state. We plot the probability density of the most localized eigenstate at each value of $t_{nnn}$ (Fig. S6I-L). In Fig. S4M, we summarize this information by measuring the probability density $\psi^2(R)$ of these localized eigenstates as a function of distance from the vacancy. As $t_{nnn}$ reaches a value of 3.25eV, the state becomes delocalized. Nevertheless, a $t_{nnn}$ value much higher than what was experimentally measured in the material ($t_{nnn}$=1) in ref. 46 still supports exponentially localized BICs.

Additional Results on Defect Dimers:

We investigate the behaviour of defect dimers in contrast to lone defects via spectroscopic imaging. In Fig. S7 we take spectroscopic maps in an area where there is a lone defect and a defect dimer. S7B-D show bonding, lone defect, and antibonding bound states in the continuum for the four-lobed bound state. S7F-H show the same for the circular bound state. In Fig. S8, we show that this dimerization behaviour requires close proximity of the defect. STM topography of a region with defect dimer or with two defects separated by an additional lattice constant can be found in Fig. S8A and B respectively. Spectroscopic maps in panels S8C-H show that defects separated by two lattice constants only host the two lone defect states, whereas those separated by a single lattice constant host dimer state. This is consistent with a tightly localized bound state in continuum.

In Fig. S9, we identify the 4 lobed state as the −1 eigenstate and the circular state as the +i eigenstate. For each of these states, we can flip the phases of all the orbitals to form degenerate eigenstates. We label degenerate states of opposite phases as (0) and (π) eigenstates. We see in Fig. S9, that the combination of two states with a π-phase relationship always yields the correct nodal structure for the lower energy states in experiment. The corresponding "antibonding" state is formed by the combination two vacancy eigenstates with the same relative phase. As such, we can conclude that the higher energy "antibonding" states must have an unfavourable interaction in the phase relationship of neighbouring vacancy eigenstates.

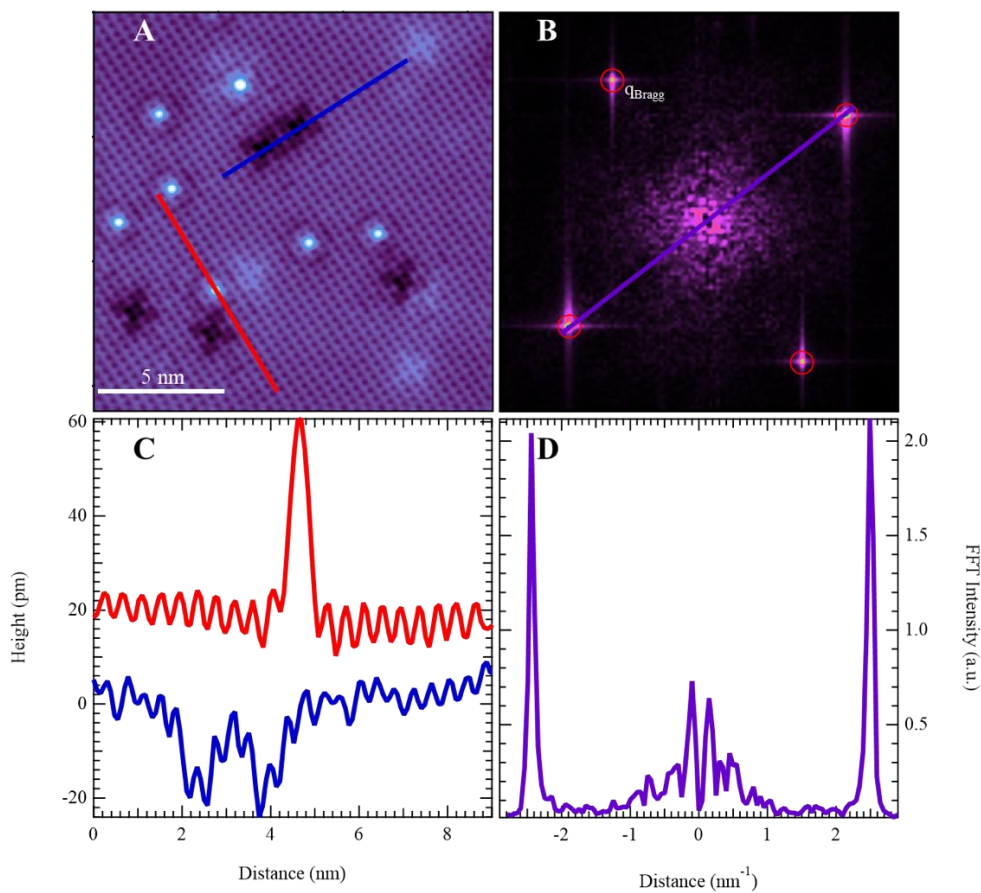

**Fig. S1. Iodine Lattice Constant and Defects**

**A.** STM topography at negative bias as shown in Fig. 1A of the main text. **B.** 2D Fast-Fourier transform of the image in A, showing Bragg wavevectors corresponding to the lattice of iodine atoms at the surface. **C.** Measurements across the blue and red lines in panel A showing the height variations due to bright and dark defects. The dark defect is centered at the I atoms which correspond to the Pd location in the PdAl plane. Bright defects are on the Al site **D**. Measurement of the Bragg wavevectors shown in B. The distance corresponds to 2.467 nm$^{-1}$ or a lattice constant of 0.405 nm.

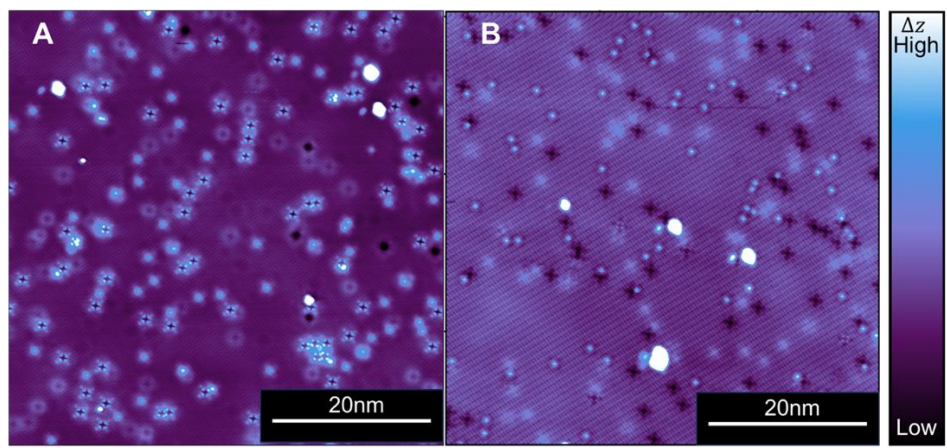

**Fig. S2. Defect Density**

STM topography at positive (**A**) and negative (**B**) sample bias over different 55×55 nm² areas. All defects are present in approximately the same density of ~70 defects in these 3025nm² areas, corresponding to a defect density of about $2.3\times10^{12}\,\text{cm}^{-2}$

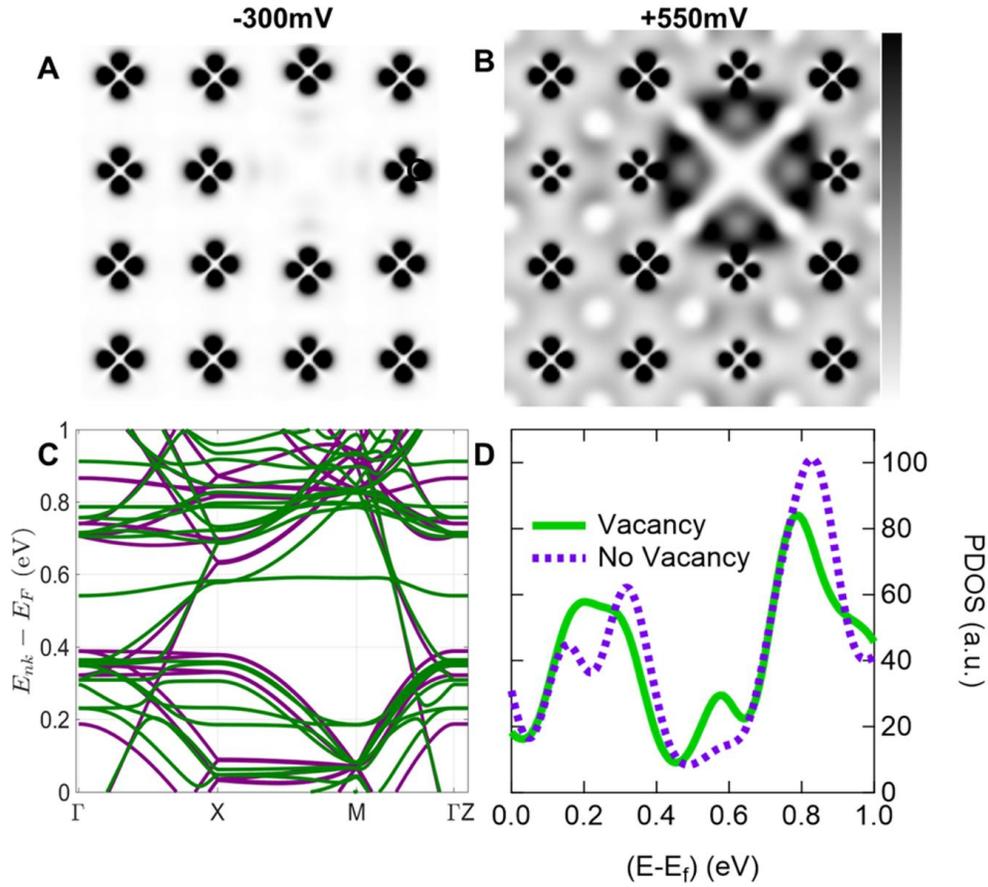

**Fig. S3. DFT Simulations**

**A, B.** Simulated STM topography of $Pd_5AlI_2$ at opposite bias polarities. A 2D slice in the PdAl plane is taken to highlight the localized density on Pd orbitals and presence of a four-lobed resonance at positive bias. **C.** DFT Band structure of $Pd_5AlI_2$ with (green) and without (purple) a Pd vacancy. The new flat band at 0.55-0.6 eV in the presence of a defect support the strongly localized nature of the observed defect state. **D.** DFT calculated partial density-of-states as a function of energy on and away from the Pd vacancy in a 4×4 supercell of $Pd_5AlI_2$. A localized defect state at 0.55-0.6 eV is evident in the spectrum.

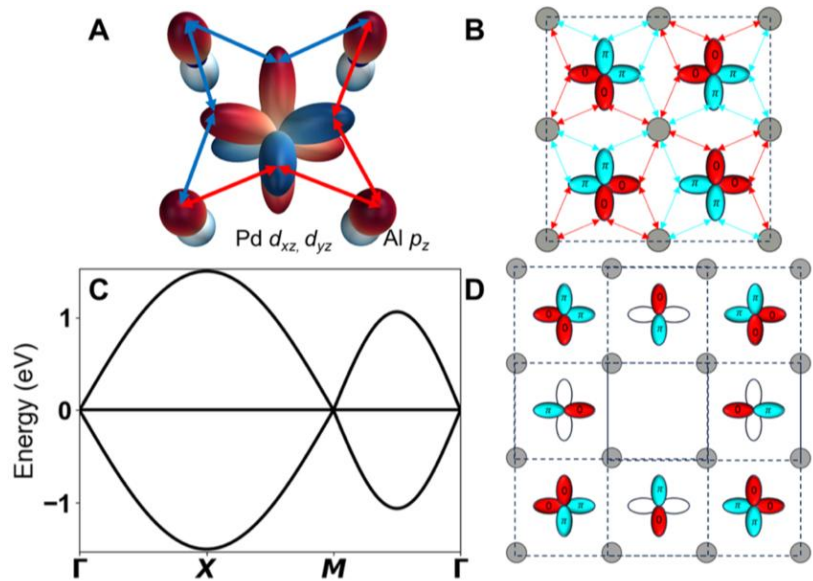

**Fig. S4. Tight-Binding Model and Compact Localized States**

**A.** Representation of the Pd *d*-orbitals and Al *p*-orbitals in the PdAl plane. The relative phase of the orbitals is indicated by red and blue colours. Blue hopping arrows are drawn between same phase lobes and red arrows between opposite phase lobes. **B.** Compact localized state in the decorated checkerboard model. Destructive interference of electron hopping from Pd to Al leads to frustrated hopping and a flat band. **C.** Band structure of the PdAl checkerboard lattice. **D.** A Pd vacancy affects the 4 compact localized states that it is a part of. The 4 CLS that are affected are combined to create the BIC, and there are 4 possible phase relationships between these 4 CLS due to 4-fold rotational ($C_4$) symmetry. These phase relationships are then captured by the +/−1 and +/−i states.

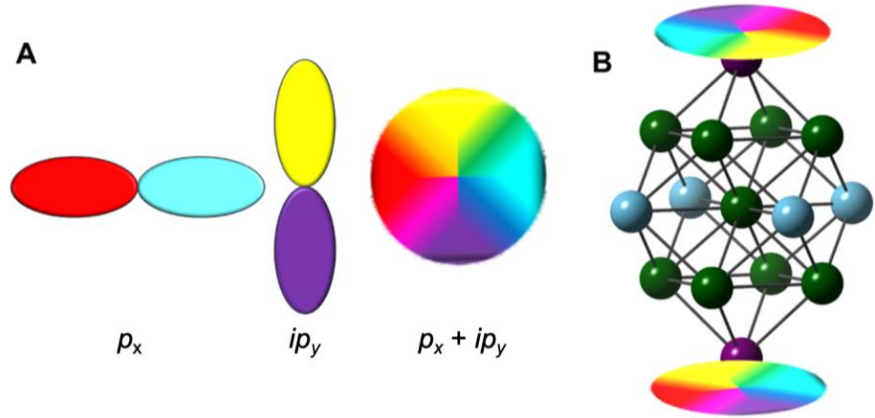

**Fig. S5. $\sigma_y$ on-site potential from local chemical potential shift in iodine atoms**

**A.** Representation of $p_x$, $p_y$ and $p_x+ip_y$ orbitals. **B.** Representation of the $Pd_5AlI_2$ lattice with a Pd vacancy in the PdAl plane and an additional on-site potential from $p_x+/-ip_y$ orbitals from the iodine atoms. This is equivalent to applying a Pauli matrix transformation of the form $m^*\sigma_y$ to the Pd $d_{xz}$ and $d_{yz}$ orbitals surrounding the defect.

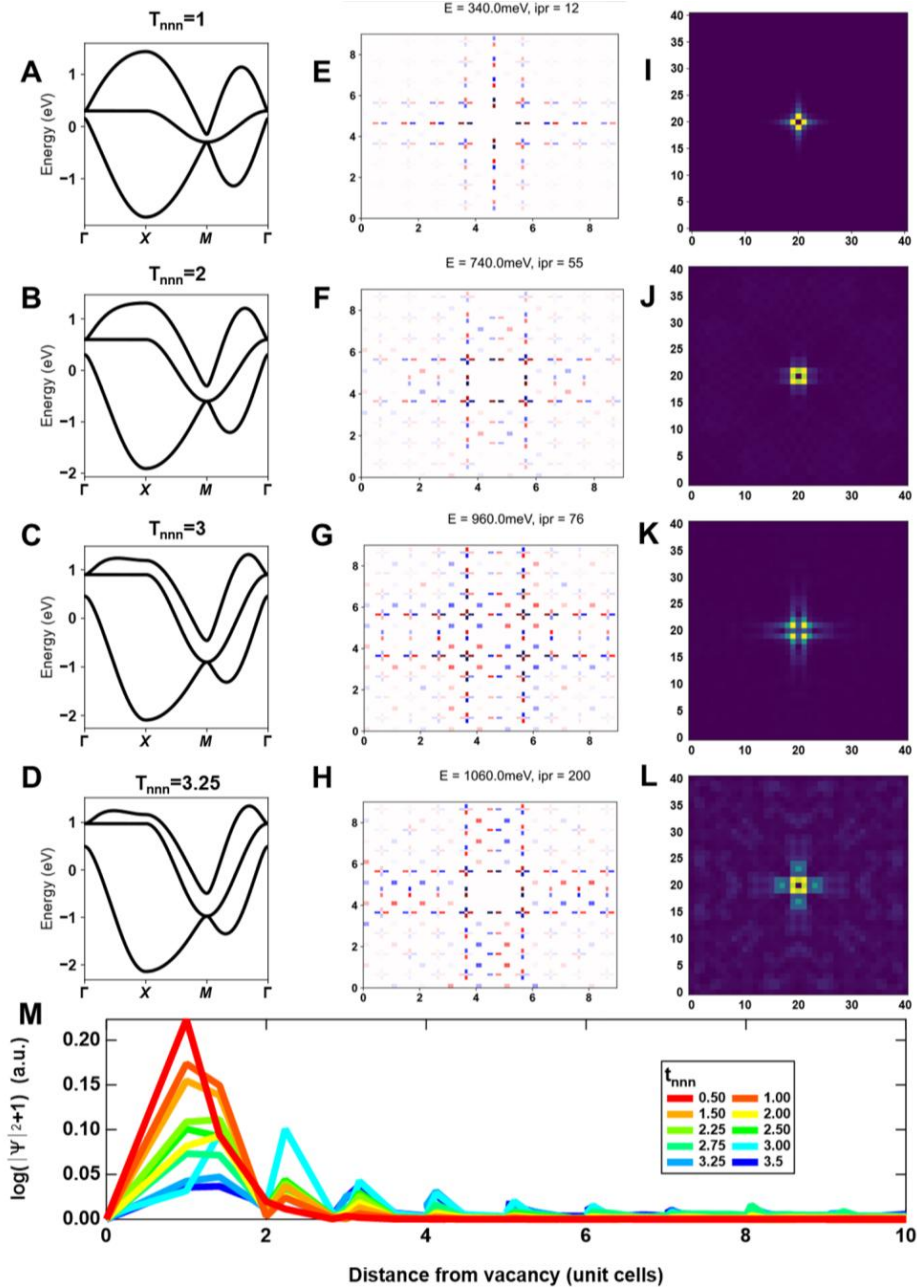

**Fig. S6. Effect of Band Dispersion on Localized Defect State**

**A-D.** Computed band structures of the PdAl 3-band model with increasing next-nearest-neighbour hopping ($t_{nnn}$). **E-H.** Calculated eigenfunctions of the lower energy localized state of a 41×41 supercell. We show the 9×9 supercell around the vacancy location. As $t_{nnn}$ is increased, the localized BIC becomes delocalized. **I-L.** Probability density ($\psi^2$) of the localized eigenstate across the whole simulated supercell. **M.** Probability density of distance from the vacancy location for different values of $t_{nnn}$. The linear relationship on a log scale shows the exponential localization of the state. The state is localized to within two unit cells from the vacancy until the next-nearest-neighbour hopping is high enough to destroy the lattice model. At $t_{nnn}=1$, which corresponds to a realistic representation of the material's band structure, the state is exponentially localized.

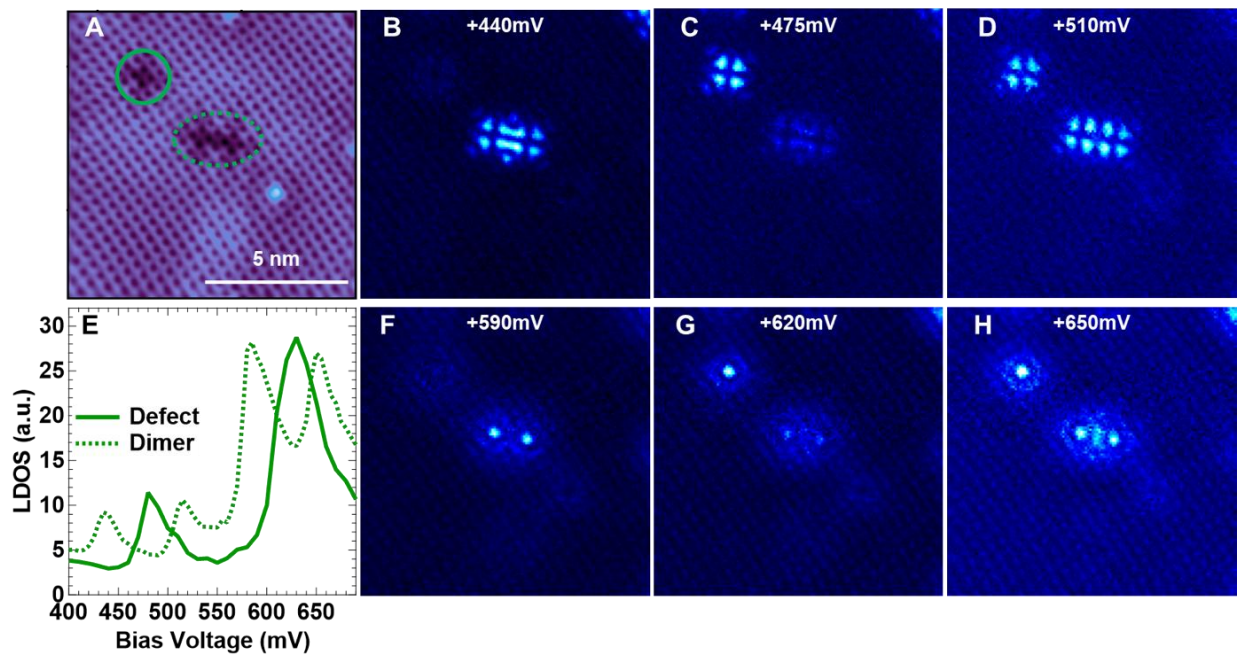

**Fig. S7. Defect Dimers Compared to Lone Defects**

**A.** STM topography of a region containing a lone dark defect and a defect dimer. **B, F.** Bonding states of the defect dimer, which lie at lower energies than the lone defect state. **C, G.** Lone defect states at energies consistent with those shown in Fig. 2 of the main text. **D, H.** Anti-bonding states of the defect dimer which lie at higher energies than the lone defect state. **E.** Average tunneling spectroscopy acquired on a defect dimer and lone defect region, showing splitting of the defect states into bonding and anti-bonding dimer states. All data were acquired with feedback setpoint of −400 mV, 50 pA and lock-in modulation of 5 mV, 1127 Hz

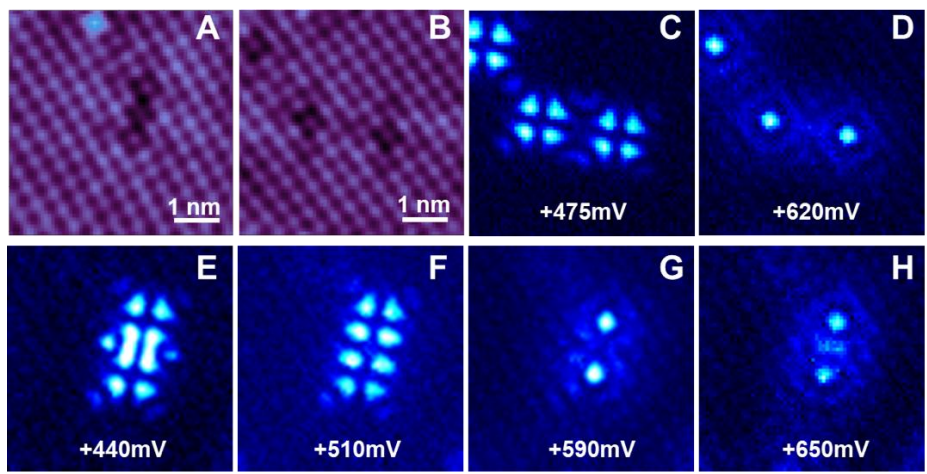

**Fig. S8. Defect Dimerization as a Function of Inter-Defect Distance**

**A.** STM topography of a region containing a defect dimer **B.** STM topography of a region containing two dark defects separated by two atomic lattice constants. **C, D.** Spectroscopic maps taken in region B showing the lone defect states. **E-H.** Spectroscopic maps takin in region A showing defect dimer states. All data were acquired with feedback setpoint of −400 mV, 50pA and lock-in modulation of 5 mV, 1127 Hz.

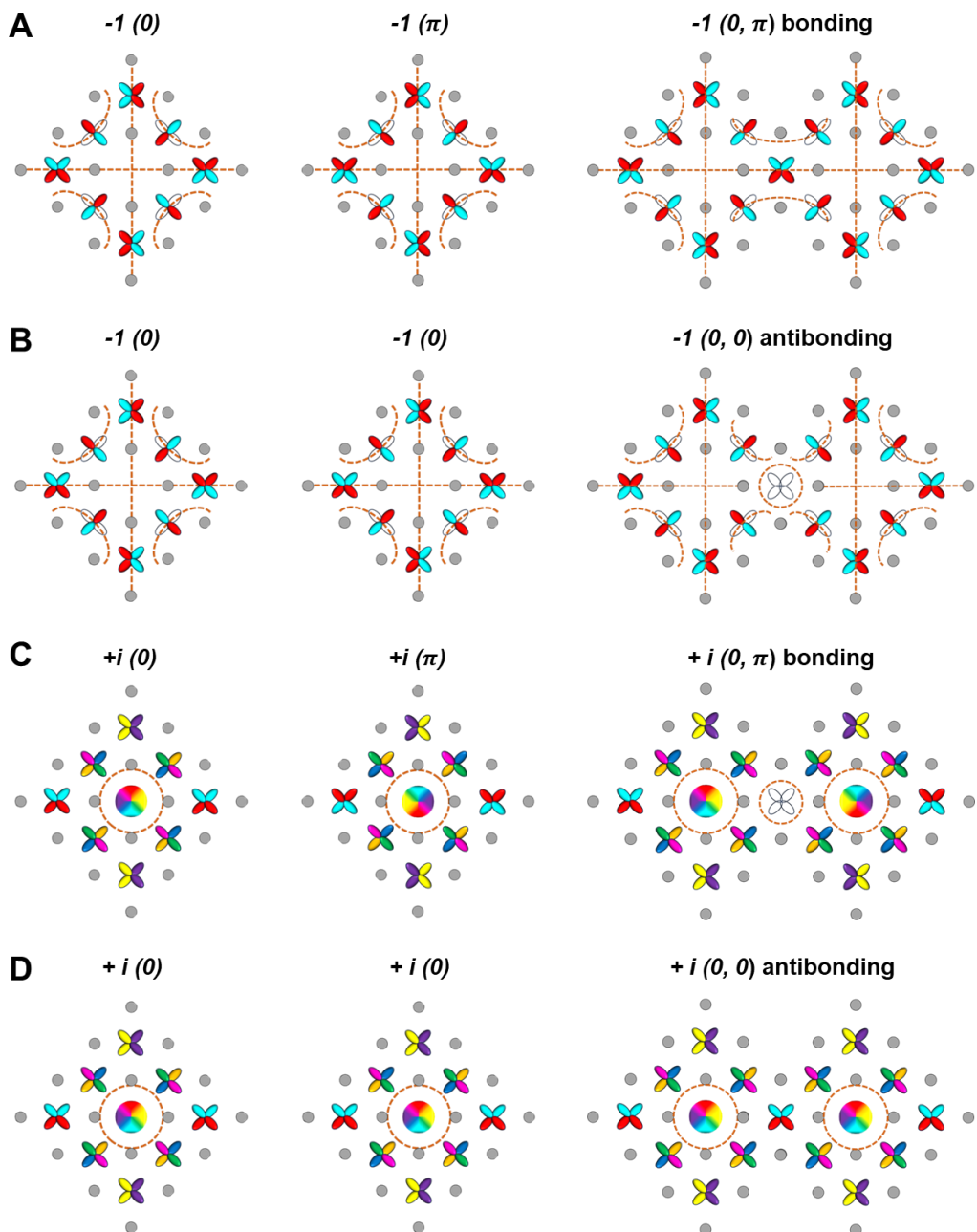

**Fig. S9. Linear Combinations of Vacancy Eigenstates form Dimer States**

**A, B.** Linear combinations of the four-lobed eigenstate and resulting nodal structure. **C, D.** Linear combinations of the circular eigenstate and resulting nodal structure. Two opposite phase states combine to form the lower energy (bonding) state. The combination of same phase states is higher in energy (antibonding).